\documentclass[aps,prl,reprint, superscriptaddress]{revtex4-1}
\usepackage{graphicx,lineno}
\begin{document}
\title{Constraints on an Annihilation Signal from a Core of Constant Dark Matter Density around the Milky Way Center with H.E.S.S.}
\author{H.E.S.S. Collaboration}
\noaffiliation

\author{A.~Abramowski}
\affiliation{Universit\"at Hamburg, Institut f\"ur Experimentalphysik, Luruper Chaussee 149, D 22761 Hamburg, Germany}

\author{F.~Aharonian}
\affiliation{Max-Planck-Institut f\"ur Kernphysik, P.O. Box 103980, D 69029 Heidelberg, Germany}
\affiliation{Dublin Institute for Advanced Studies, 31 Fitzwilliam Place, Dublin 2, Ireland}
\affiliation{National Academy of Sciences of the Republic of Armenia,  Marshall Baghramian Avenue, 24, 0019 Yerevan, Republic of Armenia}

\author{F.~Ait Benkhali}
\affiliation{Max-Planck-Institut f\"ur Kernphysik, P.O. Box 103980, D 69029 Heidelberg, Germany}

\author{A.G.~Akhperjanian}
\affiliation{National Academy of Sciences of the Republic of Armenia,  Marshall Baghramian Avenue, 24, 0019 Yerevan, Republic of Armenia}
\affiliation{Yerevan Physics Institute, 2 Alikhanian Brothers St., 375036 Yerevan, Armenia}

\author{E.O.~Ang\"uner} 
\affiliation{Institut f\"ur Physik, Humboldt-Universit\"at zu Berlin, Newtonstr. 15, D 12489 Berlin, Germany}

\author{M.~Backes} 
\affiliation{University of Namibia, Department of Physics, Private Bag 13301, Windhoek, Namibia}

\author{S.~Balenderan}
\affiliation{University of Durham, Department of Physics, South Road, Durham DH1 3LE, U.K.}

\author{A.~Balzer}
\affiliation{GRAPPA, Anton Pannekoek Institute for Astronomy, University of Amsterdam,  Science Park 904, 1098 XH Amsterdam, The Netherlands}

\author{A.~Barnacka}
\affiliation{Obserwatorium Astronomiczne, Uniwersytet Jagiello{\'n}ski, ul. Orla 171, 30-244 Krak{\'o}w, Poland}
\affiliation{now at Harvard-Smithsonian Center for Astrophysics,  60 Garden St, MS-20, Cambridge, MA 02138, USA}

% ...

\begin{abstract}
An annihilation signal of dark matter is searched for from the central region of the Milky Way. 
Data acquired in dedicated ON/OFF observations of the Galactic center region with H.E.S.S. are analyzed for this purpose. No significant signal is found in a total of $\sim 9$ h of ON/OFF observations.
Upper limits on the velocity averaged cross section, $\langle\sigma v \rangle$, for the annihilation of dark matter particles with masses in the range of $\sim 300$ GeV to $\sim 10$ TeV are derived. 
In contrast to previous constraints derived from observations of the Galactic center region, the constraints that are derived here apply also under the assumption of a central core of constant dark matter density around the center 
of the Galaxy. Values of $\langle\sigma v \rangle$ that are larger than $3\cdot 10^{-24}\:\mathrm{cm^3/s}$ are excluded for dark matter particles with masses between $\sim 1$ and $\sim 4$ TeV at $95\%$ CL if the radius of the central 
dark matter density core does not exceed $500$ pc. 
This is the strongest constraint that is derived on $\langle\sigma v\rangle$ for annihilating TeV mass dark matter without the assumption of a centrally cusped dark matter density distribution in the search region.
\end{abstract}

\maketitle

\section{Introduction}
\begin{figure}{}
\centering
\includegraphics*[scale=0.42]{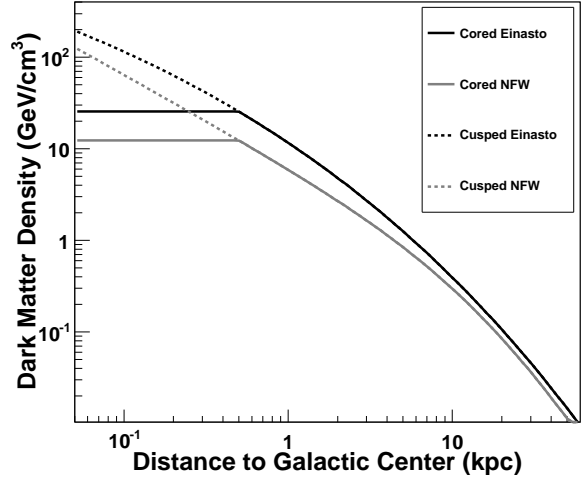}
\caption{Dark matter density as a function of the distance to the Galactic center. The parametrizations for the cusped Einasto and NFW profiles are taken from \cite{pieri}. The cored Einasto and NFW density profiles follow the 
respective cusped profiles at distances to the Galactic center that are larger than the core radius of $500$ pc.}
\label{dm_density}
\end{figure}

The formation of the large scale structure of the universe as well as the dynamics of galaxy clusters and individual galaxies strongly suggest the presence of dark matter on the respective length scale \cite{bertone}. 
Many extensions of the standard model of particle physics predict a stable particle without electromagnetic coupling whose presence can account for the missing mass that is apparent in astrophysical environments \cite{bertone}. 
The annihilation of dark matter particles is expected to produce photons with energies up to the mass of the dark matter particles \cite{tasitsiomi}. The detection of $\gamma$-rays 
from a given direction can thus indirectly probe the presence of dark matter particles along the corresponding line of sight.\\
The central region of the Milky Way is of particular interest for indirect searches for annihilating dark matter because the squared dark matter density integrated over the line of sight towards the target region (i.e. the astrophysical or J-factor) 
is expected to be large \cite{ulio}. The J-factor for observations of the 
Galactic center region depends strongly on the dark matter density distribution within the Milky Way. 
Simulations of the dynamics of the dark matter content of galaxies predict to universal dark matter density distributions. 
Towards the center of the galaxies, the influence of baryons on the distribution of dark matter is not yet resolved. 
The formation of pronounced density cusps towards the center of galaxies (\cite{gnedin}, \cite{sadeghian}) and, more recently, 
the prediction that the dark matter density in the central few hundred pc is almost constant \cite{governato}, \cite{governato2}, \cite{kuhlen} have been discussed. 
The latter prediction of an almost constant dark matter density in the central region of the Galaxy is considered in this work.\\
The current strongest constraints on the velocity averaged cross section for the self-annihilation of dark matter particles with masses in the range of $\sim 400$ GeV to $\sim 10$ TeV result from a search for an extended emission of 
$\gamma$-rays in the central region of the Milky Way with H.E.S.S. \cite{daniil}. 
However, the constraints apply only if the dark matter density distribution in the central $\sim 500$ pc of the Milky Way is cusped (see dotted lines in Fig. \ref{dm_density}). An alternative search for the annihilation of 
dark matter particles is presented in this paper and strong constraints on $\langle\sigma v\rangle$ are derived without the assumption of a dark matter density profile that is cusped in the central $500$ pc of the Milky Way.

\section{The High Energy Stereoscopic System}
The High Energy Stereoscopic System (H.E.S.S.) is an array of imaging atmospheric Cherenkov telescopes (IACT) in the Namibian Khomas Highland. IACTs detect the Cherenkov light emitted by electromagnetic showers that are induced when primary 
$\gamma$-rays interact with air nuclei in the Earth's atmosphere. Charged cosmic rays also induce showers in the Earth's atmosphere and constitute background for the IACT detection of $\gamma$-rays. Cosmic ray background events that cannot be suppressed 
during the analysis of H.E.S.S. data (see \cite{crab}) are typically treated with a background subtraction technique \cite{background}. The background subtraction relies on the definition of a signal region for which 
a background region is constructed. The construction of the background region must be performed such that the ratio of the acceptance for background events in the signal and the background region is known from instrumental characteristics. The definition of the background region enables a comparison between the 
number of events that are detected in the signal region and the number of background events that are expected in the signal region. 
The acceptance for background events of H.E.S.S. is in general strongly influenced by atmospheric conditions, the pointing zenith angle and the night sky background in the observed field of 
view. See \cite{crab} 
for more information on the H.E.S.S. experiment.

\section{ON/OFF Observations of the Galactic Center Region with H.E.S.S.}
\begin{figure}{}
\centering
\includegraphics*[scale=0.4]{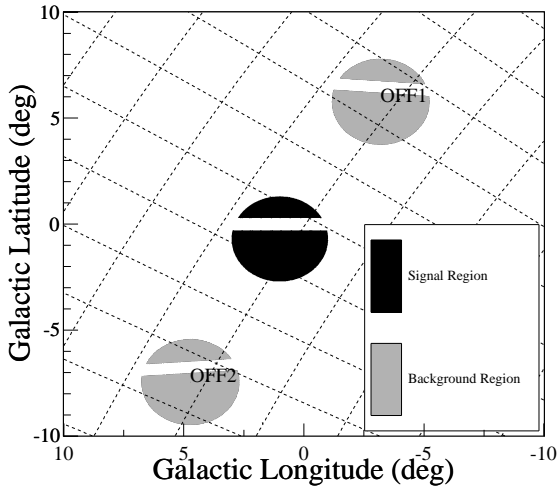}
\caption{The signal region close to the Galactic center and the two background regions with a symmetric right ascension offset of $\pm 35$ min to the signal region. 
A right ascension - declination coordinate grid is overlaid. The exclusion of the Galactic plane ($|b|<0.3^\circ$) in the signal region and the regions with a $\pm 35$ min offset in right ascension to 
$|b|<0.3^\circ$ in the background regions are visible.}
\label{onoff_pointing}
\end{figure}

\begin{figure}{}
\centering
\includegraphics*[scale=0.47]{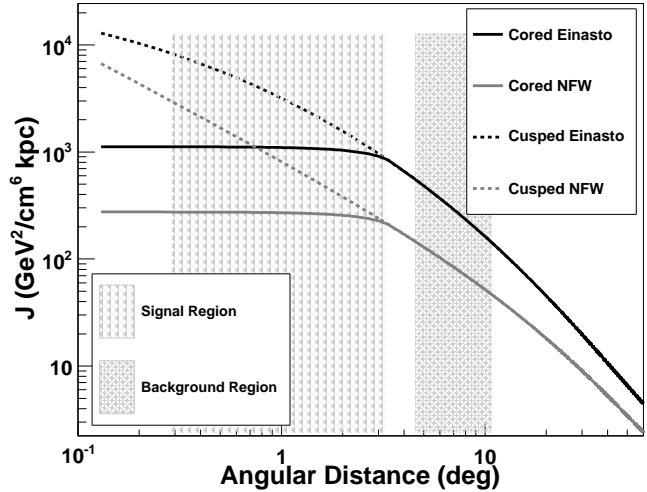}
\caption{Line of sight integral over the squared dark matter density as a function of the angular distance between the direction of the line of sight and the direction of the Galactic center. A $500$ pc core radius is assumed for the cored 
dark matter density profiles. Overlaid is the range of angular distances to the Galactic center covered by the signal and background regions of the OFF1/ON/OFF2 datasets.}
\label{los_integral}
\end{figure}

The ON/OFF observation mode (see also \cite{kosack}, \cite{background}) refers in this study to a special observation strategy where a background 
(OFF1) region, the signal (ON) region and another background (OFF2) region are observed consecutively for $33$ min each. Figure \ref{onoff_pointing} shows the observed regions in galactic coordinates. 
The signal region has a radius of $2^\circ$ and centers at $l=1^\circ$, $b=-0.7^\circ$ in galactic coordinates or $\alpha=267.7^\circ$, $\delta=-28.4^\circ$ in equatorial coordinates (J2000). The centers of the two background regions 
have a symmetric offset of $\pm35$ min in right ascension to the signal region center. 
The two minute difference between the right ascension offset between the signal and background regions and the observation length allows for a transition time between the 
observations. The ON/OFF 
observation pattern allows the equalization of the azimuth and zenith angles that are covered by array pointings in each of the observations. 
Differences in the acceptance for background events which result from differences in the zenith angle array pointing range 
can thus be neglected. The time difference of 35 min between the observations is a compromise between the demand for small atmospheric changes (i.e. small time differences) and a large offset in right ascension (i.e. large time differences). 
Two background regions are observed, to better control residual imbalances in the acceptance for background events between the observations.\\
Figure \ref{los_integral} shows the J-factor for a given line of sight as a function of the angular distance, $\theta$, 
between the directions of the line of sight and the Galactic center. The J-factor is proportional to the 
expected number of dark matter annihilation events in the respective direction. 
The $\theta$ angle ranges that are covered by the signal and background regions in the OFF1/ON/OFF2 observations are indicated in Fig. \ref{los_integral}. It is concluded from this 
figure that the expected 
number of dark matter annihilation events is larger in the signal than in the background regions when the radius of the core of constant dark matter density around the Galactic center is $500$ pc or less. 
This is a clear advantage of the ON/OFF method when compared to the background subtraction technique that is applied in \cite{daniil} which relies on the simultaneous observation of the Galactic center 
region and a background region in the same finite H.E.S.S. field of view with $\sim 2^\circ$ radius.\\
The application of standard quality criteria for H.E.S.S. data \cite{crab} and the additional requirement for compatible instrumental and atmospheric conditions within an OFF1/ON/OFF2 observation result in a total of 
six OFF1/ON/OFF2 datasets. All datasets were taken within one week in 2010 with the H.E.S.S. I array of four identical IACTs. 
The total dead-time corrected observation time for each of the three observed regions is $3.05$ h. The mean zenith angle of the array pointing for the datasets is $12^\circ$. 

\section{Data Analysis}
The image cleaning (see \cite{crab}) low and high pixel intensity thresholds for the data are chosen to be 7 pe (photo electrons) and 10 pe. Using the observed distribution of pixel intensities in cosmic ray events, it was checked that these image cleaning cut
criteria eliminate effects due to differences in sky brightness between the observed regions. 
Standard Hillas criteria \cite{crab} for the selection of $\gamma$-ray events are applied to the data. The thresholds used for image cleaning lead to an energy threshold of $290$ GeV.  
Only events with reconstructed directions within the central $2^\circ$ angular distance around the pointing position of each observation are 
considered to account for the truncation of $\gamma$-ray images near the edges of the H.E.S.S. field of view. The Galactic plane ($|b|<0.3^\circ$) is excluded from the analysis to avoid the detection of $\gamma$-rays from astrophysical sources (e.g. the Galactic center source HESSJ1745-290, \cite{sgrA}) without relation to dark matter 
annihilation. The exclusion region is shifted by the respective pointing position 
offset in right ascension into the two background regions to equalize the acceptance in the signal and background regions (see Fig. \ref{onoff_pointing}). To rule out the detection of $\gamma$-rays from astrophysical sources, the considered data with the chosen exclusion regions are analyzed with the ring background \cite{background} method and a correlation radius of 0.1 deg prior to the ON/OFF analysis. 
The resulting skymaps of the three observed regions show no indication for a significant excess. 
It is concluded from the analysis with the ring background method that the chosen exclusion regions are sufficient to exclude astrophysical sources of gamma rays for the ON/OFF analysis.\\
The mean exposure ratio, $\alpha=0.5$, for the ON/OFF data analysis is the ratio of the livetimes for the observation of the signal and background regions \cite{background}. 
However, imbalances in the acceptance for background events between the signal region and the two background regions lead to a systematic error, $\sigma_\alpha$, on the exposure ratio. 
A conservative estimate for the relative systematic error on the exposure ratio, $\sigma_\alpha/\alpha = 2\%$, is derived. This estimate results from a comparison of the number of events which pass $\gamma$-ray event selection criteria in the two background regions.

\section{Results}
\begin{figure}{}
\centering
\includegraphics*[scale=0.47]{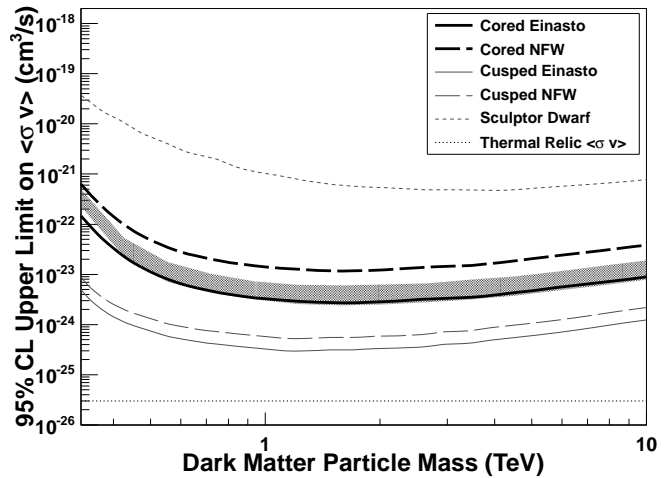}
\caption{Upper limits on the velocity averaged dark matter self annihilation cross section as a function of the dark matter particle mass. The upper limits for the cored Einasto and NFW density profiles hold for a core radius of $500$ pc and the annihilation of dark matter particles into light quarks (\cite{tasitsiomi}). The filled 
area around the upper limit curve for the cored Einasto dark matter profile shows the $\pm 1\sigma$ variations around the upper limit that is expected for this dark matter density profile when no annihilation signal is detected. The derived upper limit is stronger than the expected upper limit due to the negative significance of the measured excess. 
For comparison, the velocity averaged annihilation cross section of a 
thermal relic dark matter particle is shown. Additionally shown are the upper limits that are derived in \cite{daniil} for cusped Einasto and NFW profiles as well as the upper limit that is derived in \cite{sculptor} for a cored dark matter density 
distribution around the Sculptor dwarf galaxy.}
\label{upperLimit}
\end{figure}

A total of $N_\mathrm{ON} = 24268 $ signal and $N_\mathrm{OFF} = 49028 $ background events are measured that pass standard Hillas criteria \cite{crab} for the selection of $\gamma$-ray events. The total $\gamma$-ray signal $s$ has a statistical significance of $-0.5\sigma$. The statistical significance is calculated with the log-likelihood ratio test statistic 
as described in \cite{roostat} with the likelihood function (see also \cite{hugh}) 
\begin{equation}
L=\mathrm{P}(N_\mathrm{ON},\hat{\alpha} b + s)\mathrm{P}(N_\mathrm{OFF},b)\mathrm{G}(\hat{\alpha}, \alpha, \sigma_\alpha)\:\mathrm{.}
\label{likelihood}
\end{equation}
Here, P and G represent the Poisson and Gaussian distributions. 
The parameters $b$ (mean number of background events) and $\hat{\alpha}$ (exposure ratio with mean $\alpha$) 
are treated as nuisance parameters. For comparison, the significance of the $\gamma$-ray event excess as calculated with Eq. (17) in \cite{lima} without consideration of the 
systematic error on the exposure ratio is $-1.3\sigma$. 
Since no significant $\gamma$-ray signal is measured, an upper limit on the integrated $\gamma$-ray signal for energies ranging from the instrumental energy threshold to a maximum energy 
$\hat{E}$ is derived. For the calculation of the upper limit, the likelihood function that is given by eq. \ref{likelihood} is analyzed with the method described in \cite{roostat}. 
The upper limit on the energy integrated signal translates (see e.g. \cite{cta_dm}) into
an upper limit on the velocity averaged dark matter self annihilation cross section, $\langle\sigma v \rangle(M)$, for a dark matter particle with mass $M=\hat{E}$. 
The variation of the instrumental response with the zenith and azimuth angles of the array pointing and within the field of view is accounted for in the analysis. 
The consideration of the $2\%$ relative systematic error on the exposure ratio increases the upper limit on $\langle\sigma v\rangle$ by a factor of $\sim 3$. 
Upper limits on $\langle\sigma v\rangle$ are presented in Fig. \ref{upperLimit} for Einasto and NFW dark matter density profiles with a $500$ pc radius core of constant dark matter density around the Galactic center. 
The parameters for the NFW and Einasto density profiles are taken from \cite{pieri}. 
The derived upper limits on $\langle\sigma v\rangle$ hold for the $\gamma$-ray energy spectrum that is expected from the self annihilation of dark matter particles into light quarks (see \cite{tasitsiomi}, the same spectrum is assumed in \cite{daniil}). 
For an Einasto dark matter profile that is cored in the inner $500$ pc around the Galactic center, 
values of $\langle\sigma v \rangle \sim 3\cdot 10^{-24}\:\mathrm{cm^3/s}$ or larger are excluded for dark matter particle masses in between $\sim 1$ to $\sim 4$ TeV at $95\%$ CL. 
The upper limits on $\langle\sigma v\rangle$ that are derived for an Einasto dark matter 
density distribution with a core radius of $500$ pc are the most constraining exclusions that are derived for TeV mass dark matter without the assumption of a centrally cusped dark matter density distribution in the search region. However, these limits are one order of magnitude less constraining than the current best limits for cusped dark matter density distributions (see Fig. \ref{upperLimit}) 
and two order of magnitudes weaker than the expectation for thermal relic dark matter (see e.g. \cite{bertone}).\\
For core radii different from $500$ pc, the upper limit on the velocity averaged dark matter self annihilation cross section scales like 
$\langle\sigma v\rangle_\mathrm{R}=(\Delta J_\mathrm{500pc}/\Delta J_\mathrm{R})\:\langle\sigma v\rangle_\mathrm{500pc}$ where 
$\Delta J$ denotes the difference between the field of view averaged astrophysical factors in the signal and background region and the subscript is equal to the core radius. 
The field of view averaged astrophysical factors in the signal and background region of the considered ON/OFF analysis for different core radii are listed in table \ref{JFacs}. The upper limits 
on $\langle\sigma v\rangle$ increase by a factor of 
2 (5) if the radius of the central core of constant dark matter density is 750 pc (1 kpc) when compared to a core radius of $500$ pc. 
\begin{table}
\centering
\begin{tabular}{l|c | c | c | r}
$R$ (kpc) & $J^\mathrm{Einasto}_\mathrm{ON}$ & $J^\mathrm{Einasto}_\mathrm{OFF}$ & $J^\mathrm{NFW}_\mathrm{ON}$ & $J^\mathrm{NFW}_\mathrm{OFF}$ \\
\hline
$0$ & $2167$ & $268$ & $559$ & $78$ \\
$0.5$ & $1036$ & $268$ & $256$ & $78$ \\
$0.75$ & $636$ & $268$ & $165$ & $78$ \\
$1$ & $426$ & $255$ & $117$ & $75$ \\
$2$ & $138$ & $126$ & $46$ & $43$ \\
\end{tabular}
\caption{Field of view averaged astrophysical factors for the signal (subscript ON) and for the livetime weighted average of the two background regions (subscript OFF). 
The values are in units of $\mathrm{GeV^2\:cm^{-6}\:kpc}$ and are tabled 
for Einasto and NFW profiles as a function of the radius ($R$) of the central dark matter core.}
\label{JFacs}
\end{table}

\section{Summary}
A search for a signal from annihilating dark matter around the Galactic center was performed. For this purpose, data that were acquired in dedicated ON/OFF observations of the Galactic center region with H.E.S.S. were analyzed. No significant signal was found. 
The employed observation technique enabled the derivation of upper limits on $\langle\sigma v\rangle$ that are significantly more conservative in respect to the distribution of dark matter in the Galactic center region than previous constraints. 
In particular, the constraints apply also under the assumption of a core of constant dark matter density around the Galactic center. 
If the dark matter density in the central 500 pc around the Galactic center is constant and follows outside of the core radius an Einasto profile, 
values of $\langle\sigma v\rangle$ that are larger than $3\cdot 10^{-24}\:\mathrm{cm^3/s}$ were excluded for dark matter particle masses between $\sim 1$ and $\sim 4$ TeV at $95\%$ CL. 
This is currently the best constraint on $\langle\sigma v\rangle$ that has been derived without the assumption of a centrally cusped dark matter density distribution in the search region.

\section{Acknowledgments}
The support of the Namibian authorities and of the University of Namibia in facilitating the construction and operation of H.E.S.S. is gratefully acknowledged, as is the support of the German Ministry of Education and Research (BMBF), the 
Max Planck Society, the French Ministry of Research, the CNRS-IN2P3 and the Astroparticle Interdisciplinary Programme of the CNRS, the U.K. Particle Physics and Astronomy Research Council (PPARC), the IPNP of the Charles University, the 
South African Department of Science and Technology and National Research Foundation, and by the University of Namibia. We appreciate the excellent work of the technical support staff in Berlin, Durham, Hamburg, Heidelberg, Palaiseau, Paris, Saclay, 
and in Namibia in the construction and operation of the equipment.

\end{document}